\documentclass[12pt]{article}\usepackage[hyperfootnotes=false]{hyperref}
   \usepackage{epsfig}
      \usepackage{amsmath}
      \usepackage{amssymb}
      \usepackage{caption}

\usepackage{amsmath}

\usepackage{hyperref} 
\usepackage{setspace}
\pdfoutput=1
  \usepackage{graphicx}
  \usepackage{color}
 \newcommand{\be}{\begin{equation}}
\newcommand{\bea}{\begin{eqnarray}}
\newcommand{\eea}{\end{eqnarray}}
\newcommand{\beq}{\begin{equation}}
 \newcommand{\ee}{\end{equation}}

\def\nref#1{(\ref{#1})}

\begin{document}
  \renewcommand{\theequation}{\thesection.\arabic{equation}}

\begin{titlepage}
  \bigskip

  \bigskip\bigskip\bigskip\bigskip

  \bigskip

\centerline{\Large \bf {Bulk locality from modular flow}}

    \bigskip

  \begin{center}

 \bf { Thomas Faulkner$^1$, Aitor Lewkowycz$^2$}
  \bigskip \rm
\bigskip

\small{
 $^1$ { \it University of Illinois at Urbana-Champaign, \\
1110 West Green Street, Urbana, IL 61801, USA\\ }
 \smallskip
 \smallskip 
$^2${\it  Stanford Institute for Theoretical Physics, Department of Physics,\\ Stanford University, Stanford, CA 94305, USA}}
\smallskip

\vspace{1cm}
  \end{center}

  \bigskip\bigskip

 \bigskip\bigskip
  \begin{abstract}
We study the reconstruction of bulk operators in the entanglement wedge in terms of low energy operators localized in the respective boundary region. To leading order in $N$, the dual boundary operators are constructed from the modular flow of single trace operators in the boundary subregion.
The appearance of modular evolved boundary operators can be understood due to the equality between bulk and boundary modular flows and explicit formulas for bulk operators can be found with a complete understanding of the action of bulk modular flow, a difficult but in principle solvable task.

We also obtain an expression when the bulk operator is located on the Ryu-Takayanagi surface  which only depends on the bulk to boundary correlator and does not require the explicit use of bulk modular flow. This expression generalizes the geodesic operator/OPE block dictionary to general states and boundary regions.

 \medskip
  \noindent
  \end{abstract}

  \end{titlepage}

  \tableofcontents

  \section{Introduction}
  
Bulk locality, operators commuting at spacelike separation, is a striking feature of holographic CFTs. In the  $N \rightarrow \infty$ limit, low energy fields satisfy the free wave equation in a fixed gravitational background.  
The extrapolate dictionary \cite{BDHM} identifies bulk fields close to the boundary with local CFT operators: scalar fields $\Phi$ with a given mass are dual to single traces ${\cal O}$ with a fixed conformal weight, $\Delta$. In \cite{BDHM,Hamilton:2006az,HKLLint}, it was observed that the extrapolate dictionary together with the free wave equation determines the local bulk fields in terms of boundary operators:
 \begin{equation}
 \Phi(X)=\int d^d x f_{\Delta}(X|x) {\cal O}(x)+O(1/N) \label{HKLL}
 \end{equation}
  where the boundary integration region consists of all points that are space like separated from $\Phi(X)$. 
At leading order in $1/N$ bulk locality is then simply a consequence of large-$N$ factorization and the specific two point function  of $\mathcal{O}$ determined in this background. $1/N$ corrections can be systematically included by solving the wave equation including interactions. 
True holography enters when we realize that $\mathcal{O}(x)$ can be written in terms
of Heisenberg operators on a constant boundary time slice $\Sigma$ by explicitly evolving the operators
with the boundary Hamiltonian. Then $\Phi(X)$ encodes bulk locality 
in a highly non-trivial, yet necessarily approximate, way.

On the other hand the emergence of bulk locality is somewhat obscured by the fact
\eqref{HKLL} is a completely non-local mapping: a bulk field at some given point depends on all boundary operators which are spacelike separated from $X$. This is particularly disturbing when trying to relate the notions of bulk and boundary locality, where for example \eqref{HKLL} does not smoothly reduce
to the extrapolate dictionary as $X$ limits close to the boundary. Similarly, subregion subregion duality \cite{Czech:2012bh,Headrick:2014cta},  an attempt to relate the information contained in  boundary sub-regions to that of a dual bulk sub-region, is not usefully constrained by \eqref{HKLL}. 

In  \cite{error}, a generalization
of \eqref{HKLL} has been postulated which is better suited for the purpose of having a more local mapping between the bulk and the boundary. They proposed that bulk fields in a certain bulk subregion $r$ can be written purely in terms of (Heisenberg) CFT operators in a given boundary subregion $R$ where we will take $r,R$ to be spacelike regions of a bulk Cauchy slice $\Sigma$
such that $R$ is the intersection of $r$ with the boundary Cauchy slice, $\partial \Sigma$.  A natural mapping of subregions $r(R)$ is suggested by the holographic entanglement entropy formula \cite{Ryu:2006bv,Hubeny:2007xt,Faulkner:2013ana}:
 \begin{equation}
 S_{EE}(R)=\frac{A_{ext}(\partial r)}{4 G_N}+S_{bulk}(r)
 \end{equation}
 which tells us that the entanglement entropy of the boundary subregion $R$ is given by the area of the extremal Ryu-Takayanagi (RT) surface $\partial r$ anchored to $\partial R$ at the boundary \footnote{We are abusing notation since technically $\partial r$ also includes $R$, but it is hopefully clear that by $\partial r$ we just mean the RT surface.},  plus the  entanglement entropy of bulk QFT reduced to $r$.  More generally it is expected that bulk fields in the \emph{entanglement wedge}  $D(r)$, the domain of dependence of $r$, can be reconstructed
 in this way. 
  
A piece of evidence in favor of the entanglement wedge reconstruction proposal is based on 
the following special case. For the vacuum state in a CFT and $R$ a ball shaped region,  one can go to AdS-Rindler coordinates which cover $D(r)$ and use Rindler mode functions to write the bulk operator in $r$ in terms of the $R$ operators 

 \begin{equation}
 \Phi(X \in D(r))=\int_{D(R)} dx f^{\mathrm{Rindler}}_{\Delta}(X |x ) {\cal O}(x) \label{Rindler}
 \end{equation}
where $D(R)$ is the boundary causal domain of $R$, which in this case is simply a double light cone.
Again, with Hamiltonian evolution, we can write this as a non-local operator acting at $R$.
This is the simplest expression that one could have hoped for, however it is easy to see that such a simple expression can't be correct if one considers more general regions and states. This can be traced to the fact that more generally the entanglement wedge contains a spacetime subregion which is entirely space-like separated from $D(R)$ (the so called causal shadow region of \cite{Headrick:2014cta} ). This means that bulk operators in that region would commute with all the local operators in $D(R)$. Then a reconstruction formula analogous to \eqref{Rindler} would imply they trivially commute with one another which is inconsistent. 
 
 In  \cite{Jafferis:2015del}, it was proposed that the simplest expression that would take the previous complication into account should read:
 \begin{equation}
  \Phi(X_r)=\int_{R} dx_R \int ds f^R_{\Delta,s}(X_r|x_R) {\cal O}_s(x)\,, \qquad {\cal O}_s(x_R)=\rho_R^{-i s/2 \pi}{\cal O}(x_R)\rho_R^{i s/2 \pi} \label{HKLLJLMS}
\end{equation}  
As we will explain later, this conjugation by the density matrix is a natural operation in the field theory called modular flow. The operators ${\cal O}_s(x_R)$ are non-local and can't all commute with $\Phi(X_r)$. If $R$ is a sphere in the vacuum, this expression reduces to the Rindler expression \nref{Rindler}. 

The modular hamiltonian is defined as the logarithm of the density matrix. In theories with a holographic dual, one can think of the modular hamiltonian as an operator in bulk perturbation theory and it is given by \cite{Jafferis:2014lza,Jafferis:2015del}:
\begin{equation}
K_R=\frac{\hat{A}(\partial r)}{4 G_N}+K_{bulk,r}+O(G_N)
\end{equation}
This expression implies that the commutator of bulk operators (which are spacelike separated to $\partial r$) with the boundary modular hamiltonian is equal to the commutator with the bulk modular hamiltonian. So, an operator in $r$ commutes with the modular hamiltonian of $\bar{R}$. This property lead to the conjecture \nref{HKLLJLMS} and was argued in \cite{Dong:2016eik}  to be equivalent to quantum correctability. 

In this paper, we will derive the expression \nref{HKLLJLMS} and write a formula for the smearing function which only depends on boundary information and is in principle computable. In the limit where the bulk operator lives in $\partial r$, this formula simplifies significantly, allowing us to compute complicated boundary quantities (which depend on the modular hamiltonian) in terms of simple bulk calculations. The formula we would like to advertise relates the operators on the RT surface $\partial r$ to the modular average of the boundary operator:
\begin{equation}
\int_{-\infty}^{\infty} d s \rho_R^{-i s/2 \pi}{\cal O}(x)\rho_R^{i s/2 \pi}  =4\pi \int_{\partial r} dY_{RT} \langle \Phi(Y_{RT}) {\cal O}(x) \rangle \Phi(Y_{RT}) 
\end{equation}

The outline of the paper is as follows. We start in section $2$ by introducing various properties of modular hamiltonians and modular flows. In section $3$, we derive and explore the expression \nref{HKLLJLMS}. Section $4$ deals with entanglement wedge reconstruction when the bulk operator sits at the RT surface, which is much simpler. In section $5$, we comment on state dependence and the inclusion of interactions . We conclude with some closing thoughts in section $6$ .

\section{Modular evolution, fourier space and free fields}

Given our state of interest, $\rho$, and its associated algebra of operator ${\cal A}$, the modular hamiltonian $2 \pi K =-\log \rho$ generates an automorphism: it sends operators in the algebra to operators in the algebra. This is called modular flow \cite{Haag:1992hx} \footnote{Note that our $2 \pi s=-s_{Haag}$ in order to identify $s$ with Rindler time in the local case. }:
\begin{equation}
A \in {\cal A}_R \rightarrow A_s=e^{i K_R s} A e^{-i K_R s} \in {\cal A}_R
\end{equation}

This modular flow is interesting for various reasons. Formally, understanding its properties was important to develop Tomita-Takesaki theory, which made type III algebras tractable. In some very symmetric cases, such as half-space in the vacuum (or a sphere in the vacuum of a CFT), modular flow is generated by (local) Hamiltonian evolution \cite{Bisognano:1976za,Hislop:1981uh,Casini:2011kv} 
: $K_R=\int_R d\Sigma^{\mu} \xi^{\nu} T_{\mu \nu}$.

Given that in general this flow is non-local it will be useful to illustrate some of its properties by going to modular fourier space, a basis where the action of the modular hamiltonian is simple:  
\begin{eqnarray}
A_{w}= \int_{-\infty}^{\infty} ds e^{-i s w} e^{i K_R s} A e^{-i K_R s} \,,\qquad [K_R,A_{w}]= w A_w
\end{eqnarray}
From now on, we are going to focus on a pure state and we will consider the density matrices of subregions in that state. If we consider a subregion $R$, purity implies that $(K_{R}-K_{\bar R}) |\Psi\rangle =0$. From this, it follows that

\begin{equation}
2\langle A_w B_{w'} \rangle=\delta(w+w') \int_{-\infty}^{\infty} ds e^{-i w s} \langle A e^{-i (K_R-K_{\bar R}) s} B \rangle=\delta(w+w') \langle A_w B \rangle=\delta(w+w') \langle A B_{-w} \rangle
\end{equation}
where $A,B$ are two operators in the algebra of bounded operators in $R$. 

Furthermore, these operators satisfy the KMS conditions ($\beta={2 \pi}$) with respect to the modular flow since correlators are analytic in the strip $0<Im(s) < 2 \pi $:
\begin{eqnarray}
\langle A_s B \rangle=\langle B A_{s+2 \pi i} \rangle \,, \qquad \langle A_w B \rangle=e^{-2 \pi w} \langle B A_w \rangle \nonumber \\
 \rightarrow \langle A_w B \rangle=n_w\langle [B,A_w] \rangle \,,\qquad n_w \equiv \frac{1}{e^{2 \pi w}-1}
\end{eqnarray}

The main observation is that the KMS condition allows us to express the commutator at finite frequency in terms of the correlator. This is an important property: the commutator of any operator and any other operator at finite modular frequency will be zero iff they are not correlated. 

In the following, we are going to consider local operators, ${\cal O}(x)$ in region $R$. We will denote their modular frequency two point function $G_w$:
\begin{equation}
\langle {\cal O}_{w}(x) {\cal O}_{-w'}(x') \rangle= \delta(w-w') G_w(x,x')=n_w \langle [{\cal O}_{-w}(x),{\cal O}_{w'}(x')] \rangle
\end{equation}
By definition, correlators in modular time are time translation invariant. If we had a translation invariant state, we could go to momentum space and canonically normalize the modes. In general, $G_w(x,y)$ is not translation invariant and one can't simplify the modes further. 

As a side note, the generator of the full modular flow $U(t)=e^{-i(K_R-K_{\bar R})t}$ is well defined in the continuum limit: it acts like the usual modular flow when acting in an operator in $R$ or $\bar{R}$ but it is also an good operator, which doesn't depend on details near the entangling surface $\partial R$. In this way, sometimes is useful to think of the previous correlators in terms of:
\begin{equation}
\langle A U(t) B \rangle
\end{equation}
which is equivalent to the previous discussion when $A,B$ are in the same subregion, but it is well defined more generally. 

\subsection{Gaussian states in free field theory}

Gaussian states in free theories have been explored extensively in the literature (see for example \cite{Casini:2009sr} and references therein). Since all correlators are fixed by the two point function, the density matrix will be gaussian and the modular hamiltonian bilinear. We will be focusing on scalar fields, the operators in the algebra of the region ${\cal R}$ \footnote{This is a field theoretical discussion, but in the next section, when we apply this to holography, this will be a subregion of the bulk. } are $\lbrace \Phi(X),\Pi(Y); \forall x \in {\cal R} \rbrace$  and the respective modular hamiltonian will be:
\begin{align}
\label{modhamphi}
&K_{\cal R} =\int_{\cal R}  dX \int_{\cal R} dY \vec{\Phi}(X).{\cal K}(X,Y)  .\vec{\Phi}(Y) \\ \qquad \vec{\Phi}(X) &\equiv (\Phi(X),\Pi(X)) \,,  \qquad \Pi(X)=n^{\mu} \partial_{\mu} \Phi(X) 
\nonumber
\end{align}
where $n^\mu$ is the unit term to the Cauchy slice containing $\mathcal{R}$. 

In general,  ${\cal K}$ is unknown unless the system is specially symmetric: one can write it formally in terms of matrix inversions and log's of correlators, but these are hard to obtain in the continuum limit (and only understandable in a distributional sense) although one can obtain it numerically using a lattice.  Free fields are interesting in the holographic setting because to leading order in $1/N$, the bulk quantum theory consists of free fields. Subleading orders introduce weak interactions that can be included perturbatively in the previous discussion: for example if we have a $\Phi^3$ interaction, there will be a contribution to the modular hamiltonian which will be trilinear on the fields. 

The modular fourier modes will then be a linear combination of $\Phi,\Pi$:
\begin{equation}
\Phi_w(X)=\int_{\cal R} dY (c_w(X,Y) \Pi(Y)-n^{\mu}\partial_{\mu} c_w(X,Y)  \Phi(Y) )
\end{equation}
where the coefficients are determined by the canonical commutation relations: $ c_w(X,Y)=-i (e^{2 \pi w}-1) \langle \Phi_w(X) \Phi(Y) \rangle$.

These modular frequency modes can also be used as a basis of operators. This is how the entanglement entropy is computed in \cite{Casini:2009sr}. The idea is that instead of labeling the modes by $(d-1)$ coordinates $x$, we can label them by $w$ and $(d-2)$ coordinates $X_S$, which correspond to a codimension $2$ surface $S \subset R$. If we think of the $c_w(X_S,Y)$ as a matrix $C_{Y';Y} \propto G_{w}(X_S,Y)$ with index $Y'=(w,X_S)$, then different $Y,Y'$ have different values and $C_{Y';Y}$ is always different from zero (we don't expect the correlators to vanish). Furthermore, no linear combination of $\Phi$,$\Pi$ can give zero, so this matrix is invertible. The invertibility of this matrix guarantees that $\Phi_w(X_S)$ is a good basis, ie there is a Bogoliubov transformation that sends $\lbrace \Phi(Y),\Pi(Y) \rbrace \rightarrow \lbrace \Phi_w(X_S), \Phi^{\dagger}_w(X_S)\rbrace$. This basis is useful because the modular hamiltonian is simpler and the entropy can be computed easily.  

In terms of the modes $\Phi_w(X_S)$, the modular hamiltonian will read:
\begin{equation}
K_R=\int_{w>0} dw \int_S dX_S \int_S dY_S w n_w {\cal K}_w(X_S,Y_S) \Phi_w(X_S) \Phi_{-w}(Y_S) \label{bulkKw}
\end{equation}
This hamiltonian is slightly simpler compared with \eqref{modhamphi}: it has one less integral. The commutation relations determine ${\cal K}$ in terms of $G_w$:
\begin{align}
[K_R,\Phi_{w}(X_S)]= w \Phi_{w}(X_S)&= w \int_S dY_S dZ_S {\cal K}_{w}(X_S,Y_S) G_{w}(Y_S,Z_S)  \Phi_{w}(Z_S) \nonumber \\
 &\hspace{-2cm} \rightarrow  \int_S dY_S {\cal K}_{w}(X_S,Y_S) G_{w}(Y_S,Z_S)=\delta(X_S-Z_S) \label{modhamscalar}
\end{align}
where we should think of ${\cal K}$, the modular hamiltonian kernel as a distribution, which can be thought as the matrix inverse of $G_{w}$.   Of course,  this discussion seems very formal because in order to define the modular frequency modes, one has to know the modular hamiltonian in the first place. However, as a pratical tool, it seems like it might be simpler to compute $G_w(X,Y)$ than evaluating the modular hamiltonian \nref{modhamphi} in terms of the $\Phi,\Pi$ directly. The reason is that this two point function has a very particular analytic structure: it picks up a phase when going around the entangling surface $\partial {{\cal R}}$ . Solving the wave equation with these branch cut boundary condition seems like an easier way to handle the problem and this is how \cite{Casini:2005zv} dealt with the entanglement entropy massive scalar field and a spherical region. This approach to the entanglement entropy of free fields seems to be practically useful also for less generic regions  \cite{Casini:private}.

\section{A smearing function in the entanglement wedge}

The extrapolate dictionary  relates boundary and bulk operators in the boundary of $AdS$:
\begin{equation}
\lim_{z \rightarrow 0} z^{\Delta} \Phi(x,z) = {\cal O}(x)
\end{equation}
In the future, we will denote this as $\Phi(x,z=0)={\cal O}(x)$.
Given that the bulk field is a free field, one can use the equations of motion to move this field to the bulk, at the expense of including more boundary operators:
\begin{equation}
\Phi(X)=\int dx f(X|x) {\cal O}(x)
\end{equation}
One normally presents $f(X|y)$ as the Green's function for the scalar field with support at spacelike separation. However, $f(X|x)$ can be  understood by formally inverting the bulk-to-boundary correlator\footnote{This formalism doesn't make the support at space-like separation explicit, but this can be accounted for by inverting the boundary two point function only around the points which are space-like separated from the bulk point. This can be done explicitly in fourier space if one restricts the sum over frequencies, see \cite{Banerjee:2016mhh} for example. }:
\begin{align}
\langle \Phi(X) {\cal O}(y) \rangle&=\int dx f(X|x) \langle {\cal O}(x) {\cal O}(y) \rangle \\
f(X|x)=\int dy \langle \Phi(X) {\cal O}(y) \rangle \tilde{K}(y,x) \,, &   \quad \int dz\tilde{K}(y,z) \langle {\cal O}(z) {\cal O}(w)  \rangle=\delta(y-w)
\end{align}
which should be understood as a distribution. 

For example, if the system under consideration has spatial translation symmetry and time translation symmetry (as in the case of the vacuum, finite temperature or Rindler), these kernels are quite simple in fourier space \cite{Papadodimas:2012aq}. We can fourier transform all bulk directions but one, which we will denote generally as $z$. Given that the two point function is diagonal in momentum space, $\tilde{K}_{w,k}=G_{w,k}^{-1}$. $\Phi(z,x)$ is normally expanded in terms of canonically normalized annihilation and creation modes $\Phi_{w,k}(z)=\psi_{w,k}(z)a_{w,k}$,  the extrapolate dictionary relates $a_{w,k}={\cal O}_{w,k}G_{w,k}^{-1/2}$, so $f_{w,k}(z)=\psi_{w,k}(z) G^{-1/2}_{w,k}$. This summarizes the explicitly known cases of HKLL and, as emphasized in  \cite{Papadodimas:2012aq,Morrison:2014jha}, makes it clear that the kernel $f(X|x)$ is defined only as a distribution.

\subsection{Entanglement wedge reconstruction}

In \cite{Jafferis:2015del}, the modular hamiltonian of holographic theories was studied to leading order in $1/N$. They concluded that the commutator of a bulk field in the entanglement wedge with the modular hamiltonian was given by the commutator with the bulk modular hamiltonian:
\begin{equation}
[\Phi(X_R),K_{R,bdy}]=[\Phi(X_R),K_{R,bulk}]+O(1/N)
\end{equation}

That is, in holographic theories, modular flow is bulk modular flow\footnote{At this point one could complain about the fact that   \cite{Jafferis:2015del} only proved this relation for $[K,\Phi]$, but not the exponentiation. However, it is easy to argue that $[K,\Phi]$ is a low energy operator, and it keeps us within the ``code subspace'' of low energy bulk perturbation theory. To make this argument one should think of computing these commutators by analytically continuing the (derivatives of the) Renyi entropies: it is clear that a low energy operator in the original geometry will be low energy in the replicated geometry. We thank Don Marolf for asking this question.}:
\begin{equation} \label{modJLMS}
e^{-i K_R s} \Phi(X_R) e^{i K_R s}=e^{-i K_{R,bulk} s} \Phi(X_R) e^{i K_{R,bulk} s}
\end{equation}
This equation is suggestive for subregion reconstruction and  \cite{Jafferis:2015del} conjectured that there is a smearing function in terms of these modular evolved modes:
\begin{equation}
\Phi(X_R)=\int ds \int_{R} dx f_s(X|s) {\cal O}_s(x)=\int dw \int_{R} dx f_w(X|x) {\cal O}_w(x) \label{hkllmod}
\end{equation} 
In this section we are going to derive and explore this formula.  To do this, we will modular evolve the extrapolate dictionary $\Phi(x,z=0)={\cal O}(x)$. If we apply \nref{modJLMS}  to the extrapolate dictionary, we get:
\begin{equation}
\Phi_s(x,z=0)={\cal O}_s(x) \rightarrow \Phi_w(x,z=0)={\cal O}_w(x) \label{Extra}
\end{equation}

This new version of the extrapolate dictionary seems quite shocking at first: it maps a non-local boundary operator to a smeared bulk operator with support in all the entanglement wedge. When the modular hamiltonian is local\footnote{Because of the extrapolate dictionary, the bulk modular hamiltonian is local if and only if the boundary modular hamiltonian is local. }, this seems standard since we can go to the Schrodinger picture, but, in the Heisenberg picture, this should seem surprising at first. Only if the modular hamiltonian is local can ${\cal O}_s(x)$ be understood as a local operator at some other point in the causal domain of $R$.  This will be the key property behind entanglement wedge reconstruction.  It is important to note that, while the modular flow of a single trace will be in general a very complicated operator in the region, \nref{Extra} implies that they have a simple expression in terms of single traces at all boundary times, because one can just do global HKLL on the modular flow of the bulk operator.

\subsubsection*{A change of basis for free fields}

We can now apply the free field theory discussion of the previous section when the subregion is  ${\cal R}=r$, a space-like section of the bulk entanglement wedge. As we discussed, the expression 
\begin{equation}
\Phi_w(X_S)=c_w \int_{r} dY \left(\langle \Phi_w(X_S) \Phi(Y) \rangle \Pi(Y)- \langle \Phi_w(X_S) \Pi(Y) \rangle \Phi(Y) \right)
\end{equation}
should be thought as a Bogoliubov transformation  $\lbrace \Phi_w(X_S) \rbrace \rightarrow \lbrace \Phi(Y) \rbrace$. In order to use the extrapolate dictionary, we should consider $S$ to be the intersection of $r$ with the boundary that is we want to set $S$ to be the boundary region $R$. The extrapolate dictionary then implies that:
\begin{equation}
{\cal O}_w(x_R)=c_w \int_{r} dY \left(\langle {\cal O}_w(x_R) \Phi(Y) \rangle \Pi(Y) -\langle {\cal O}_w(x_R) \Pi(Y) \rangle \Phi(Y)\right) \label{changebasis}
\end{equation}
As we argued before, we can invert this matrix and its inverse gives the smearing function for entanglement wedge reconstruction. The key point is that, in the bulk, we have been able to write any bulk operator in the entanglement wedge in a basis where the modular extrapolate dictionary acts naturally. The fact that entanglement wedge reconstruction could be implemented by a Bogoliubov transformation was mentioned in \cite{Kim:2016ipt}, however \cite{Kim:2016ipt} didn't identify what set of modes in the entanglement wedge has a localized boundary representation.

\subsubsection*{Symmetry based argument}

Alternatively, given that the bulk and boundary modular flow are the same, one could look directly for a boundary representation of the bulk operator that satisfies the following requirements:
 \begin{equation}
 \Phi_{w}(x_R,z \rightarrow 0) ={\cal O}_w(x_R) \,, \quad [K_{ R},\Phi_{w}(X_R)]= w \Phi_{w}(X_R) \,,\quad[K_{\bar R},\Phi_{w}(X_R)]=0 ; 
 \end{equation}
The latter requirement implies that this operator is in $R$ and the other two conditions force this operator to have fixed modular frequency and have a fixed conformal dimension $\Delta$. The only single trace operators that satisfy these requirements are the operators of region $R$ with fixed modular frequency: ${\cal O}_w(x_R)$\footnote{From the generalized extrapolate dictionary, it is clear that this operator is a single trace, because it is a linear combination of bulk fields which are single traces themselves}. This way of finding a boundary representation of the bulk operator is reminiscent of  \cite{NO1}. 

To leading order, the bulk field should be expandable in single trace with fixed modular frequency. This symmetry based approach seems more useful to justify the existence of a smearing function to higher order in matter interactions (or changes in the background metric), since this would require higher traces coming from bulk interactions to appear with fixed modular frequency.

\subsection{The smearing function}

The smearing function is determined by requiring it reproduces the correct bulk-to-boundary correlator:
\begin{align} 
\Phi(X)= \int_{-\infty}^{\infty} dw\int_R dx f_{w}(X|x) {\cal O}_{w}(x) \,, \quad & f_{w}(X|x) \equiv \int_R dy  {\cal K}_{w}(x,y) \langle \Phi(X) {\cal O}_{-w}(y) \rangle \nonumber \\
&\int dy {\cal K}_{w}(x,y) \langle {\cal O}_{w}(y) {\cal O}(z) \rangle=\delta(x-z) \label{smearing}
\end{align}

  Note that $f_w(X|x)$ satisfies the free wave equation in the bulk variable and ${\cal K}_w(x,y)$ can be thought of as being defined implicitly in terms of the boundary correlators \nref{smearing} or alternatively as the bulk modular hamiltonian of \nref{bulkKw}\footnote{This expression implies that the distributional character of the smearing function is on the same footing as that of the bulk modular hamiltonian .}. All equalities are to be understood as expansions around a fixed state in the $G_N \rightarrow 0$ limit. We expect that one can introduce corrections due to interacting matter by just demanding this expression to give the right bulk-boundary-boundary three point function with $f$ satisfying the corrected wave equation in the presence of interactions. The symmetry based approach of the previous section might be more convenient for this. 
 
 One often thinks of the HKLL kernel $f_w(X|x)$ as the solution to the wave equation with certain boundary conditions in a fixed background. Our approach is a combination  of a change of basis in the bulk and the modular extrapolate dictionary. However, note that we could have also obtained $f_w(X|x)$ following the usual approach: requiring it that it has $\delta$ function support when $X \rightarrow x$, that it satisfies the wave equation and that it adquires a $e^{-2 \pi w}$ monodromy when going around RT surface (this condition appears naturally when considering the wave equation that $\langle \Phi_w(X) \Phi(Y) \rangle$ satisfies). 
    
Note that $f_w(X|x)$ is an expression purely in the boundary: ${\cal K}_w$ is the inverse of $G_{w}=\int ds e^{-i w s} \langle {\cal O}_s(x) {\cal O}(y) \rangle$ and $\langle \Phi(X) {\cal O}_w(y) \rangle$ can be computed using global HKLL and the knowledge of the boundary modular flow. Alternatively, one can compute this bulk-to-boundary modular correlator in terms of the bulk modular flow which might be easier to implement explicitly.

To make contact with \nref{hkllmod}, we can also write the explicit $f$ in modular time:
\begin{equation}
f_s(X|x)=\int_R dy \int ds' \langle \Phi(X) {\cal O}_{s-s'}(y)\rangle {\cal K}_{s}(x,y) 
\end{equation}
with ${\cal K}_{s}$ the fourier transform of ${\cal K}_w$. As was explained in \cite{Morrison:2014jha,Papadodimas:2012aq}, in the local case we don't expect $f$ to be a proper function, but a distribution, which when evaluated within correlators gives a definite answer, this is clear from our approach since the bulk modular hamiltonian, ${\cal K}_w$, is a distribution.

\subsubsection*{Different boundary representations for the same bulk operator}

As we have seen, it seems like there are different equivalent representations for the bulk fields: they can be written in terms of boundary fields smeared over different subregions. In \cite{error}, it was suggested that this is realization of quantum error correction. In more simple terms, these two boundary representations for the bulk operator have to differ by an operator that is zero within low energy correlators. As was explained in \cite{Mintun:2015qda,Freivogel:2016zsb}, in the vacuum, these operators correspond to single trace operators with space-like momenta. 

More generally, we expect that the reason why there are different representations for bulk operators has to do with the fact that $[K_R-K_{r,bulk},\Phi(X_r)]=0$ when inserted in low energy correlators. It is this approximate equality between the action of bulk and boundary modular hamiltonians that allows us to localize the field: from the global HKLL expression it is not clear whether $[K_{\bar R},\Phi(X_r)] = 0$, but since this is equal to the commutator with the bulk modular hamiltonian this is trivially zero.

\subsubsection*{The effective generalized free fields modular hamiltonian}

From the boundary point of view, the theory satisfies large $N$ factorization. To leading order in $1/N$ we can think of the single trace operators as generalized free fields (GFF). Generalized free fields are defined as fields whose commutator is a c-number. By themselves they are not very nice quantum field theories, see for example \cite{Haag:1963dh}, but they often arise in an approximate manner. GFF look like free fields but they don't satisfy any equation of motion, which means that operators at different times are independent. In holographic theories, ${\cal O}_s(x)$ is a single trace operator (because of the extrapolate dictionary) and independent of ${\cal O}(x)$, so it seems that if one wanted to define the algebra of generalized free fields constrained in a subregion\footnote{Of course, this concept doesn't make sense in the purely generalized free field theory, but only as the effective algebra of a subregion in the large $N$ limit.}, one should include all operators of the form ${\cal O}_s(x)$.

 One could construct the effective modular hamiltonian of the subregion, and because to leading order in the large $N$, we just have a gaussian theory, $K_{GFF,R}$ will be bilinear. That is, if ${\cal O}(x)$ is a generalized free field, $K_{GFF}$ is the bilinear operator that effectively (within two point functions) generates the same modular evolution as the modular hamiltonian: $\langle e^{i K_R s} {\cal O}(x) e^{-i K_R s} {\cal O}(y) \rangle=\langle e^{i K_{GFF,R} s} {\cal O}(x) e^{-i K_{GFF,R} s} {\cal O}(y) \rangle$.  This condition is enough to get a explicit expression for $K_{GFF}$:
\begin{align}
K_{GFF}& =\int dw \int_R dx dy w n_w {\cal K}_{w}(x,y) {\cal O}_{w}(x) {\cal O}_{-w}(x)  \,, \\
& \int dy {\cal K}_w(x,y) \langle {\cal O}_w(y) {\cal O}(z)\rangle=\delta(x-z)
\end{align} 
 
Given the previous section, we don't expect this to be an independent object and, of course, ${\cal K}_w(x,y)$ is just the boundary limit of the bulk modular hamiltonian \nref{bulkKw}.   One can understand the $1/N$ perturbation theory by adding higher order terms, as one does in the bulk. From this expression, one could equivalently derive the smearing function by demanding the right commutation relations with the modular hamiltonian $\Phi_w(X)=\frac{[\Phi_w(X),K_{bulk}]}{ w}=\frac{[\Phi_w(X),K_{GFF}]}{ w}$, which expresses $\Phi_w(X)$ in terms of ${\cal O}_w$ because the commutator of two single trace operators is just a c-number to leading order.

\section{The zero mode}

In the previous section, we demonstrated how one should think about entanglement wedge reconstruction if one has access to the boundary (or bulk) modular hamiltonian. The resulting formula were however not very explicit and involved formal inversions of integral kernels. 
It would be nice if one can get some more explicit expressions to work with. In this section we show how to do this in the case of the modular zero modes. 

Consider the bulk operators $\Phi_w(X)$, for small modular frequency $w \sim 0$. These should approximately commute with the modular hamiltonian. From the bulk point of view, this operator has to be roughly localized near the RT surface. In the next subsection, we will show that the previous expression \nref{changebasis} simplifies dramatically in the $w \rightarrow 0$ limit, to give:
\begin{equation}
{\cal O}_0(x)=4\pi \int_{RT} dY_{RT} \langle \Phi(Y_{RT}) {\cal O}(x) \rangle \Phi(Y_{RT}) \label{zeromode}
\end{equation}
When the modular hamiltonian is local, such as for a spherical region in a CFT, this expression generalizes the OPE block story of \cite{block,deBoer:2016pqkda,daCunha:2016crm}.
In those papers, instead of having the zero mode of a local operator on the left hand side, one smears the zero mode over the boundary region $R$. Because a sphere in the vacuum is conformally equivalent to a the hyperboloid at temperature $1/2 \pi$, they consider uniformly integrating the field over the hyperboloid, which is a particularly symmetric operator. This allows one to use symmetry
arguments to show that the result is the bulk operator integrated over the RT surface.
Our expression is consistent with this - one can explicitly check, in the hyperboloid conformal frame, that integrating \eqref{zeromode} over the boundary region $R$ gives exactly the geodesic (minimal surface) bulk operator on the right hand side.

When the modular Hamiltonian is non-local \eqref{zeromode} is a highly non-trivial and surprising generalization of this OPE block story. Note  that due to the reduced symmetry in this case there is no natural way to integrate \eqref{zeromode} over the boundary region $R$.

At this point we will address a subtlety that might have bothered some readers.
One might worry that we don't really want the operator to sit exactly on the RT surface - the operators in the algebra have to be properly smeared into the region $D(r)$. In particular we don't expect the bulk scalar field theory to have a center - so the notion of a well defined operator in the algebra commuting with the modular Hamiltonian $K_R$ and $K_{\bar{R}}$ will necessarily be approximate. Or perhaps a more constructive way to think about this is that both sides of \eqref{zeromode} should really be thought of as operator valued distributions, just as is the case for local quantum fields. So the zero mode may diverge when inserted into correlation functions. This divergence could arise either because of the $\int_{-\infty}^\infty ds$ integral defining the zero mode, or equivalently because the right hand side of \eqref{zeromode} is not property smeared over a co-dimension $0$ region in the bulk spacetime. Of course \eqref{zeromode} will also often be finite when inserted in correlation functions, so it is perfectly reasonable to talk about this operator.  We will see some of these delicate issues in the derivation below.

If the reader is not interested in the derivation, she can jump ahead to the next subsection, where we discuss the implications of \eqref{zeromode}.
 
 \subsection{Derivation of the zero mode formula}
 
The derivation we present here works from a slightly different starting point compared to our discussion in Section 3. In particular we have found several ways to derive this zero mode formula, we present the method here that we find most instructive. For a derivation of \eqref{zeromode} which starts with \eqref{smearing} see Appendix~\ref{AB}. 
 
It is convenient to consider a more general  formula to \nref{zeromode} where we take any local operator in the entanglement wedge and compute it's zero mode using the bulk modular Hamiltonian $\Phi_0(X)$.
This operator also commutes with the modular Hamiltonian so one might expect it can be reconstructed on the RT surface also.

Firstly note that since the bulk is a free theory, for Gaussian states, the resulting modular zero mode will necessarily be an integral of $\Phi$ and $\Pi$ on a fixed Cauchy slice:
\be
\Phi_0(X) = \int_\Sigma \sqrt{h} d Y  \left( f_\Pi(Y) \Pi(Y) + f_\Phi(Y) \Phi(Y) \right)
\ee
where $\Pi = n^\mu \partial_\mu \Phi$, $n^\mu$ is the unit normal to $\Sigma$ and $h$ is the induced metric on $\Sigma$.  Note that we can compute the unknown functions $f_{\Pi,\Phi}$ by commuting both sides with the field operators and using the canonical commutation relations. We find:
\be
f_\Pi(Y) = - i \left< \left[ \Phi(Y), \Phi_0(X) \right] \right>  \qquad f_\Phi = - n^\mu \partial_\mu f_\Pi
\ee
where we have used the fact that the bulk theory is quadratic to replace the c-number commutator with the expectation value. Note that this expression for $f_\Pi$ is independent of the Cauchy slice $\Sigma$ on which it is evaluated. At this point we simply have to constrain $f_\Pi$. Since the zero mode that appears in these commutators has very special properties we will easily be able to fix $f_\Pi$. For example we can apply the KMS condition:
\begin{align}
\label{modflowcomm}
f_\Pi (Y) &=  i \int_{-\infty}^{\infty} ds \left( \left< \Phi_s(X)   \Phi(Y) \right>
- \left< \Phi(Y)   \Phi_s(X) \right> \right) \\ 
&=  i \int_C ds  \left< \Phi_s(X)   \Phi(Y) \right>
\label{modflowcorr}
\end{align}
where the later contour $C$ lies along the lines ${\rm Im} s = 0, 2 \pi$ at the boundaries of the complex $s$ strip. Since via very general considerations the modular flow correlator in \eqref{modflowcorr} should be analytic in the strip we can complete the contour, assuming the
large $|s|$ behavior is sufficiently decaying, and \nref{modflowcorr} will yield $f_\Pi =0$. This will be correct for $Y$ any bulk point contained in the entanglement wedge $D(r)$. For points in $D(\bar{r})$, the original commutator vanishes \eqref{modflowcomm} because the two operators are contained in commuting algebras. However there is a question of exaclty what happens at the RT surface $\partial r$. In particular our Cauchy slice will be chosen to pass through the RT surface. Since this is where the zero mode is expected to live we must analyze $f_\Pi$ carefully here.

\begin{figure}[h!]
\begin{center}
\vspace{5mm}
\includegraphics[scale=0.7]{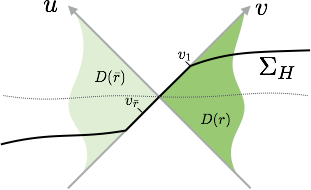}
\includegraphics[scale=0.7]{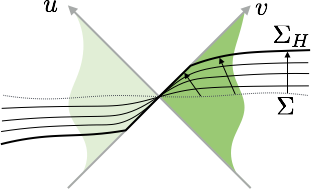}
\vspace{5mm}
\caption{(\emph{left}) We reconstruct the zero mode using the ``Cauchy slice'' $\Sigma_H$ with a small null segment close to the RT surface. (\emph{right}) We should really take a limit of space-like slices that approaches $\Sigma_H$. }
\label{sigmah}
\end{center}
\end{figure}

In order to smooth out the behaviour of $f_{\Pi}(Y)$ near the RT surface, let us approach $\partial r$ as follows. We will pick the Cauchy slice $\Sigma = \Sigma_H$ to lie along the future part of the ``horizon'' $\partial D(r)$ close to the RT surface where it will be a null light sheet. We also pick $\Sigma_H$ to continue smoothly onto a segment of the light sheet contained in $\partial D(\bar{r})$ as pictured in Figure~\ref{sigmah}. Further away from the RT surface we then take $\Sigma_H$ to smoothly match onto a spacelike surface.

Since $\Sigma_H$ has a null segment we imagine approaching it as a limit of a sequence of spacelike Cauchy slices $\Sigma \rightarrow \Sigma_H$. This however brings about some additional complications.  We now have to worry about an order of limits issue - consider a regulated version of the zero mode:
\be
\Phi_0^{(T)}(X) \equiv \int_{-T}^{T} d s e^{ i K_R s} \Phi(X) e^{- i K_R s}
\ee
where we will eventually send $T \rightarrow \infty$. The associated function $f_\Pi$  can be computed for this regulated zero mode as above. We want to first push $\Sigma$ to $\Sigma_H$ and then take the zero modes, so the appropriate order of limits we should take is:
\be
\label{order}
f_\Pi(Y_H \in \Sigma_H) = -i  \lim_{T \rightarrow \infty} \lim_{(Y \in \Sigma) \rightarrow Y_H} \int_{-T}^{T}
ds \left< \left[ \Phi(Y), \Phi_s(X) \right] \right> 
\ee
and equivalently for $n^\mu \partial_\mu f_\Pi (Y_H \in \Sigma_H)$. If we had taken the other order
of limits we would have again found $f_\Pi =0$ away from the RT surface, and although this would still likely give rise to the correct answer for $\Phi_0$, no value is gained compared to simply taking $\Sigma$ space-like. 

Keeping the order of limits \eqref{order} in mind we now  argue for a different behavior for $f_\Pi$ for points
along the null segments of $\Sigma_H$. Firstly note that $f_\Pi$ still vanishes on $\Sigma_H \cap \partial D(\bar{r})$
since the operators within $D(\bar{r})$ always commute with $\Phi_s(X)$ which means the first limit
in \eqref{order} will yield zero even for finite $T$. 

Now let us address the points on $\Sigma_H \cap \partial D(r)$.
Since we are free to act with modular flow in \eqref{order} to the left onto the operator $\Phi(Y)$ one might guess that sending $s \rightarrow T \rightarrow -\infty$ will now yield an operator that gets ``stuck'' on the RT surface and so the correlator will not decay for large $-s$. Thus we are no longer able to complete the $s$ contour as we did when $Y$ was a bulk point in $D(r)$. Let us assume this is true, and we will confirm this a posteriori.
Now the contour argument we used in \eqref{modflowcorr}  will give a non-zero contribution from the vertical piece of the contour:
\be
f_\Pi(Y) =  -\lim_{T \rightarrow - \infty}  \int_0^{2 \pi } d \theta \left< \Phi_{T-i \theta}(X) \Phi(Y)  \right>\,,
\qquad Y \in \Sigma_H \cap \partial D(r)
\ee
Note that the large $T \rightarrow + \infty$ limit will still give a correlator that vanishes sufficiently fast so we can complete the contour there. For example as we take $Y \rightarrow Y_{RT}$ the modular flow of this operator will be independent of $T$ and $\theta$ so we will have:
\be
\label{rtlimit}
\lim_{Y \rightarrow Y_{RT} } f_\Pi(Y) =- 2 \pi \left< \Phi(Y_{RT}) \Phi(X) \right>
\ee
We will now show that $f_\Pi(Y)$ is constant along null generators of the future lightsheet in $\partial D(r)$ and so it will take the value given by the limit \eqref{rtlimit} everywhere on the future part of the null generator\footnote{ This suggests that $f_\Pi$ is a highly singular function especially when any of these null geodesics leaves $D(r)$ along seams that emanate from a caustic (see \cite{Headrick:2014cta} for some nice pictures of how this happens in a relevant context). Since in this case two different null generators, with different values of $f_\Pi$ intersect.  This is to be expected since modular flow will have some singular behavior for operators close to these seams, however we can avoid these by picking $\Sigma$ to leave the lightsheet before we ever encounter any caustics. Note that $f_\Pi$ is already a distribution since it vanishes everywhere inside $D(r)$ and can be anyway regulated by picking $T$ large but finite. }.

We will only need to do this analysis in the neighborhood of the RT surface where we pick local Rindler coordinates:
\be
ds^2 = - du dv + h_{ij} d y^i d y^j + \ldots
\ee
Consider the surface $\Sigma$ defined as $u=0$ for $ v_{\bar{r}} < v < v_1$ where $v_{\bar r} < 0$ lies on the appropriate part of $\partial D(\bar{r})$ and $u = u_\star(v) < 0$ for $v > v_1$. That is $\Sigma_H$ leaves the null sheet beyond $v > v_1$. We evaluate $\Phi_0(X)$ here with what we so far know about $f_\Pi$:
\be
\Phi_0(X) = \left( f_\Pi, \Phi \right)_{\Sigma_H}
= \int d^{d-2} y \sqrt{h} \left( \int_{v_{\bar r}}^{v_1-\epsilon} d v ( f_\Pi \partial_v \Phi -  \Phi \partial_v f_\Pi )
- \int_{v_1 -\epsilon}^{v_1 + \epsilon} dv \Phi \widehat{n}^\mu \partial_\mu f_\Pi \right)
\label{finalterm}
\ee
where $( \Phi_1,\Phi_2)_{\Sigma_H}$ is the KG symplectic flux integrated over $\Sigma_H$. We
have split the contribution into two parts - the first from the null segment and the second from close to where $\Sigma_H$ leaves the null segment. For the later we have dropped $ f_\Pi \widehat{n}^\mu \partial_\mu \Phi $ since this will be smooth and thus vanish as we send $\epsilon \rightarrow 0$. All fields are evaluated at $(u=u_\star(v),v)$ where we take $u_\star(v) =0$ for $v < v_1$. 
Note that the appropriate (rescaled) normal vector is:
\be
\widehat{n} = \partial_v - u_\star'(v) \partial_u
\ee
We can evaluate the final term in \eqref{finalterm} by noting that:
\be
\label{delta}
\widehat{n}^\mu \partial_\mu f_{\Pi} = 2 (\partial_v f_\Pi)(u_\star(v),v) - \frac{d}{d v} f_\Pi(u_\star(v),v)
\ee
where the first term will be smooth as a function of $v$ so we can drop it and the later term will have a delta function since as we leave the null sheet $f_\Pi$ goes from non-zero to zero\footnote{
This delta function will be naturally resolved using the $T$ regulator with a width $=\mathcal{O}(e^{-T} )$.}. Putting this together we have:
\be
\label{phi0int}
\Phi_0(X) = - 2 \int d^{d-2} y \sqrt{h} \int_{v_{\bar r}}^{v_1-\epsilon} d v \Phi \partial_v f_\Pi(0,v)
\ee
where we have integrated by parts. The boundary term vanishes at $v = v_{\bar r}$ since $f_\Pi$ vanishes there and at $v=v_1-\epsilon$ it cancels the delta function term in \eqref{delta}.
 
Let us now imagine picking a slightly different surface $\Sigma_{H,2}$ which leaves the null segment at a slightly later null coordinate $v=v_2$. The answer should be independent of this choice, so we find that:
\be
0 = \int d^{d-2} y \sqrt{h} \int_{v_1}^{v_2} d v \Phi \partial_v f_\Pi
\ee
and since this is an operator statement the only way we can satisfy this is if $\partial_v f_\Pi =0$
for $ v>0$. 
Note that we can pick the null coordinates differently on each generator $v_{1,2}(y)$ .
We have thus shown that the equations of motion necessitate that $f_\Pi$ is constant along null generators of $\partial D(r)$. We made the argument near the RT surface using local Rindler coordinates, but that was just for simplicity and this argument should be applicable away from this coordinate patch (although our zero mode derivation does not rely on this fact.)

Across the horizon now $f_\Pi$ jumps to zero so \eqref{phi0int} gives:
\be
\Phi_0(X) = - 2 \int_{RT} dY_{RT} \Phi(Y_{RT}) \lim_{Y \rightarrow Y_{RT} }f_\Pi(Y) = 4\pi\int_{RT} dY_{RT}  \left< \Phi(Y_{RT}) \Phi(X) \right> \Phi(Y_{RT})
\ee
where we have used \eqref{rtlimit}. 
This was the claimed result. Since this derivation might seem delicate we give another derivation, making use of a different limit towards the horizon, in Appendix~\ref{AB} and find the same answer. This new derivation relies on some reasonable assumptions about modular flow for free QFTs close to the entanglement cut/RT surface.

\subsection{Bulk reconstruction from zero modes}

In the context of bulk reconstruction, we can invert the previous formula to write 
\begin{equation}
\Phi(X_{RT})=\int_R dx f_0(X_{RT}|x) {\cal O}_0(x)
\end{equation}

This formula gives a simple formula for reconstructing bulk fields in the entanglement wedge if we can foliate $r$ by RT surfaces of smaller regions. This requires inverting the bulk-to-boundary correlator over the RT surface, which is easier than having to solve the full modular flow.

\subsubsection*{More on smearing and conserved charges}

In quantum field theories, one normally works with smeared operators: $\Phi_f=\int d^d x f(x) \Phi(x)$, which have finite two point functions. These operators are normally smeared in codimension $0$ surfaces, but one could wonder if they could be defined in higher codimensional surfaces. Let's consider the case of a free scalar field, smeared around a codimension $n$ surface (see for example \cite{Bousso:2014uxa}):
\begin{equation}
\int d^{d-n} x d^{d-n} y f(x) f(y) \langle \Phi(x) \Phi(y) \rangle \propto \int \frac{d^{d-n} x}{x^{d-2}} \sim \epsilon^{2-n}
\end{equation}

So one can't smear a scalar field in a codimension $2$ surface. This means that the operator localized in the RT is not a well defined operator and it is not contained in the algebra of operators of the entanglement wedge. 

For this reason, whenever we discuss these zero modes, we should think of smearing it inside the entanglement wedge. In this way, this becomes an operator of the respective region $R$ (or $\bar{R}$) and not an operator which is seemingly in the center. Equivalently, since we are working to leading order in $1/N$, we can't resolve frequencies which are $O(1/N)$ which means that we should put some $O(N)$ cutoff in the boundary integral over modular flow. 

To some extent, this is a subtlety, but it is important because we don't expect to have a center for this simple example with bulk scalar fields. 

An exception to this analysis would be that of conserved charges. Conserved charges, such as the electric field in the RT surface or the area operator can be localized in a codimension two surface, because Gauss' law allows it. From the algebraic point of view, this is no mystery, since we expect these algebras of operators to have a center, see \cite{Casini:2013rba}. 

Understanding conserved charges is really interesting but we are not going to explore these in detail.  We expect that their boundary smearing function should simplify and it is likely that the vector, $\xi$,  with which they are contracted is also simple. For the case of gravity, we expect that graviton contribution to the linearized area operator should correspond to the zero mode of the stress tensor:
\begin{equation}
\delta A=\int_{RT}\sqrt{\gamma} \gamma^{i j} \delta h_{i j}=\int dx ds f_T(x) T_{\xi \xi}(x|s)
\end{equation}

In the electromagnetic case, we expect that:
\begin{equation}
\int_{RT} E.dS=\int dx ds f_J(x) J_{\xi}(x|s)
\end{equation}
The electric flux along the RT surface was studied before in  \cite{Hartnoll:2012ux}. For Rindler, these conserved charges were studied in \cite{deBoer:2016pqkda} and since the modular hamiltonian is local there, the $s$ integral drops out. Note that we only expect these integrals of conserved charges to be well defined without further smearing when Gauss' law applies. We don't expect this to be true for general bulk smearings over the RT surface.

\subsubsection*{Dressing}

In all our discussion, we have ignored the contribution from the area to $K_{bdy} \sim K_{bulk}$. As explained in the previous section, the zero modes shouldn't be exactly localized in the RT surface, but even if they were, they would commute with the area operator, since the area operator just generates a time translation. The only operators with which the area operator does not commute are operators in its past/future\footnote{ If we consider gravitational corrections, some other fields that would seem to commute with the boundary modular hamiltonian would be bulk operators dressed to the extremal surface. Understanding the role of these operators seems very interesting, but we leave this for the future.}.

In gravitational theories, operators can't be exactly local, they have to be defined in a gauge invariant way. This means that they can't commute with the stress tensor everywhere in the boundary because Gauss's law relates the energy in the bulk with the energy flux at infinity. One normally thinks of this in terms of gravitational dressing: the bulk operator has its own gravitational field which gives a non-trivial commutator in the boundary, see \cite{Donnelly:2015hta} for more details. 
 
From the bulk point of view, there are many ways that a bulk field could be  gravitationally dressed. In analogy with the wilson line in electromagnetism, a geodesically dressed case is often considered, where the gravitational flux is localized at a point in the boundary \cite{Heemskerk:2012np,Kabat:2013wga,Guica:2016pid,Fitzpatrick:2016mtp}, but more non-local dressings seem also natural from the boundary perspective \cite{Lewkowycz:2016ukf} .

Our zero modes (slightly smeared into $R$) commute with all the operators in $\bar{R}$ , (approximately) with the modular hamiltonian $K_R$ and within correlators they commute with all operators in the region: $\langle [A_0,B] \rangle \approx 0$, for an arbitrary operator $B \in R$. If the modular hamiltonian is local, $[H_R,A_0]=[K_R,A_0]\sim 0$ would mean that Gauss' law is properly satisfied and that the gravitational field is localized around $\partial R$. More generally, we expect that, given that the operator approximately commutes with the modular hamiltonian, this is enough to guarantee that this operator should be localized in $\partial R$. If we had a nice quantum field theory, we certainly expect that these zero modes are basically operators localized in $\partial R$, but because of large $N$, these could be  non-trivial operators. 

While we leave a more detailed analysis for the future, we expect that these zero modes are dressed to the boundary codimension $2$ region, $\partial R$. It is unclear to what extent a zero commutator with the modular hamiltonian is enough to conclude that Gauss' law is satisfied but our operators satisfy $\langle [\Phi_0,T_{\mu \nu}]\rangle=0$. We expect that, after taking care of the gravitational interactions, $[\Phi_0,T_{\mu \nu}(x_R)]=0$ inside correlators.

\subsection{Zero modes and $G_0(x,y)$}

Equation \nref{zeromode} also gives us a simple bulk method to compute $G_0(x,y)$:
\begin{equation}
G_0(x,y)=\langle {\cal O}_0(x) {\cal O}(y)\rangle=4\pi\int dY_{RT} \langle  {\cal O}(x)\Phi(Y_{RT}) \rangle \langle \Phi(Y_{RT}) {\cal O}(y) \rangle \label{G0eq}
\end{equation}

This quantity is very non-trivial to compute in the field theory, but in gravity it can be computed explicitly given the RT surface and the bulk-to-boundary correlator in a particular state. For example, if the operators have a large external dimension $N^2 \gg \Delta \gg 1$:
\begin{equation}
\frac{-1}{\Delta} \log G_0(x,y) =L(x,Y_{RT,min})+L(Y_{RT,min},y) 
\end{equation}
where $L(X,y)$ is the geodesic distance between a bulk and a boundary point and $Y_{RT,min}$ is the point in the RT surface which minimizes the previous expression.

In appendix \ref{AA}, we prove this formula explicitly in the local case, using the  methods of \cite{Faulkner:2014jva}. In the next subsection, we compute $G_0(x,y)$ for some simple holographic examples and compare it with the field theory expectations.  

\subsubsection*{Two interval example}

A simple, illustrating example is that of two intervals in two dimensions in the vacuum. Consider the $t=0$ spacelike slice and put the two intervals at $(-1,-y) \cup (y,1)$. $G^{2-int}_0(x_1,x_2)$ will just be the sum over two one interval modular correlators $G_{0}^{(u,v)}(x_1,x_2)$, where, for $\Delta=1$, $G_0^{(-L,L)}(x_1,x_2)$ is given by:
\begin{equation}
G_0^{(-L,L)}(x,_1,x_2)\mathop{=}_{\Delta=1}\frac{2 L}{(x_1-x_2)(x_1 x_2-{L^2})}\log \frac{(L-x_1)(L+x_2)}{(L+x_1)(L-x_2)}
\end{equation}
In this way, the two intervals zero mode correlators are given by:
\begin{eqnarray}
G^{2-int}_0(x_1,x_2)=G_{0}^{(-1,-y)}(x_1,x_2)+G_{0}^{(y,1)}(x_1,x_2)\,, \quad \text{if } y<y_0 \nonumber \\ 
G^{2-int}_0(x_1,x_2)=G_{0}^{(-1,1)}(x_1,x_2)+G_{0}^{(-y,y)}(x_1,x_2)\,, \quad \text{if } y>y_0  
\end{eqnarray}

\begin{figure}[h!]
\begin{center}
\vspace{5mm}
\includegraphics[scale=0.75]{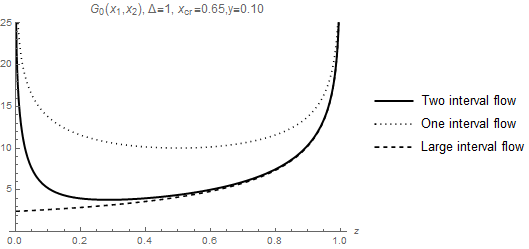}

\vspace{5mm}
\caption{$G^{2-int}_0(\frac{1-y}{2},y+z (1-y))$ as a function of $z\in (0,1)$ for $\Delta=1$ and $y=0.1$ (which corresponds to a cross ratio of $x=0.65$). The large interval flow corresponds $G_0^{(-y,y)}(x_1,x_2)$ and the one interval flow to that of $G_0^{(y,1)}(x_1,x_2)$. We see that when $x_2$ is close to the endpoints of the interval, the flow is roughly that of the smaller interval, as we expect from locality. For $z>1/2$, the modular flow is well approximated by that of the large interval. }
\label{2int}
\end{center}
\end{figure}

In Figure~\ref{2int}, we plot  $G^{2-int}(x_1,x_2)$. We can compare it with the modular flow of the large interval $(-y,y)$ and the modular flow of the single interval $(y,1)$.

\subsubsection*{Comparison with 2d free fermions}

 The only analytic example of non-local modular flow was studied in \cite{Casini:2009vk,Longo:2009mn} for $2d$ free fermions and $n$ intervals.  Given that this is a usual quantum field theory, we expect these zero modes to be roughly localized in $\partial R$. 
 
 In this case, one can compute the modular flow for two intervals, at $(-1,-y),(y,1)$ and the modular flow will be non-local but it will only mix two trajectories, one in each interval. Furthermore, it doesn't mix chiralities. The modular evolution of the chiral fermionic field operators undergoes mixing:

\begin{equation}
\Psi_s(x_1)= \cos \theta(x_1,s) \Psi(x_1(s)) +  \sin \theta(x_1,s)\Psi(x_2(s))
\end{equation}

with $x_1$ being the initial location of the operator, inside one of the intervals (of course, $\Psi(x_1,s=0)=\Psi(x_1)$). The explicit expression for $\theta(x_1,s)$ can be found in \cite{Casini:2009vk,Longo:2009mn} and $x_1(s),x_2(s)$ are the corresponding local modular trajectories , which are given explicitly by the two solutions to: 
\begin{equation}
\frac{(x_{1,2}(s)-y)(x_{1,2}(s)+1)}{(x_{1,2}(s)+y)(x_{1,2}(s)-1)}=e^{-s} \frac{(x_1-y)(x_1+1)}{(x_1+y)(x_1-1)}
\end{equation}

In \cite{Longo:2009mn} they compute the modular evolved correlator explicitly to make sure that it satifies the KMS condition
\begin{equation}
\langle \Psi_s(x) \Psi(z) \rangle=\frac{e^{-s/2}}{X e^{-s}-Z } \frac{X-Z}{x-z} \,,\quad  X(x)=\frac{(x-y)(x+1)}{(x+y)(x-1)}
\end{equation}

The zero mode correlator will then be
\begin{equation}
G_0(x,z)=\frac{1}{\sqrt{X(x)} \sqrt{Z(z)}} \frac{X(x)-Z(x)}{x-z}
\end{equation}

After integrating over modular flow the pole at $x=z$ disappears and one instead gets two branch cuts when $X(x)Z(z)=0$, that is, when $x$ or $z$ are light-like separated from the left boundaries of the intervals. This is what one might have expected for chiral fermions. 

\section{Interactions, background dependence and backreaction}

It is interesting to understand how these formulas for entanglement wedge reconstruction, which depend on the modular flow of the state under consideration, are compatible with the usual background dependent description of HKLL. Similarly it is interesting to understand how to include interactions. Of course, our expressions are valid for any state as long as we don't need to shift the entanglement wedge (and thus the algebra of operators stays the same). However if the background fields change or one considers higher orders in bulk perturbation theory, this will be no longer be true. 

\subsubsection*{Interactions} 

In the presence of $1/N$ corrections, the HKLL dictionary between bulk and boundary fields is not diagonal anymore, let us illustrate this with a $\lambda \Phi^3$ bulk interaction. In the vacuum, these corrections to HKLL read \cite{HKLLint}:
\begin{align}
\Phi(X)&=\int d^{d}x  \left [ f_{\Delta}(X|x) {\cal O}(x)+\lambda \sum_{n} f_{2 \Delta+2 n}(X|x) (O \partial^{2 n} O)(x) \right ]+... \nonumber \\
\Phi_{w,k}&=K_{w,k}^{\Delta} {\cal O}_{w,k}+\sum_n f_{w,k}^{2 \Delta+2 n} (O \partial^{2 n} O)_{w,k}+...
\end{align}
where the new term is a sum over double trace operators.
Of course, the $1/N$ expansion encodes the expression of the bulk fields in other states which are related by a shift in the one point functions. It contains more information since it reproduces higher  order correlators, which can't be understood as two point functions in a different state.

From the argument of the previous sections, we expect that our modular fourier mode expression has the same expansion:
\begin{equation}
\Phi_{w}(X_R)=\int f_w(X|x) {\cal O}_w(x)+\sum_n \int  f_{n,w}(X|x) (O \partial^{2 n} O)_w(x)+...
\end{equation} 
That is, bulk operators of fixed modular frequency can be understood in terms of a single trace operator plus double trace operators of the same frequency. Gravitational or gauge interactions can be taken into account by solving the wave equation perturbatively and will add corrections for the boundary field which include the boundary stress tensor and currents. 

In the context of zero modes, one can imagine accounting for the interactions in two different ways. One could stick to the zero mode of a single trace operator ${\cal O}_0$ and thus the bulk field integrated in the RT surface would get $1/N$ corrections, or keep fixed the bulk field $\Phi$ integrated in the RT surface and study what boundary zero mode it corresponds to. While the first approach was explored in the Rindler case in  \cite{deBoer:2016pqkda}, we expect the second approach to be simpler, since it is aligned with the usual ideas of HKLL, but we reserve a more careful analysis to the future. 

\subsubsection*{Backreaction}

The discussion of the previous sections referred to matter interactions or changes in the background fields, but didn't include backreaction. In this work, we have been using heavily the fact that the bulk and boundary modular flow are equivalent.  However, if we want to account for higher $G_N$ corrections, the results of \cite{Jafferis:2015del} will receive order $G_N$ corrections and the two modular flows won't be exactly the same  \cite{DongLewk}. Despite  this, \cite{Patricketal} have shown that when difference between bulk and boundary relative entropies is small (as in the case of $G_N$ corrections), one can still reconstruct the bulk operator. Even if the position of the surface changes by a small amount, the modular hamiltonian will change everywhere \cite{Faulkner:2016mzt}, but the previous arguments seems to suggest that one can still work in perturbation theory.    In this way, we expect then that \cite{Patricketal} together with \cite{DongLewk} allow our story to be generalizable in the presence of backreaction. 

\subsubsection*{Background dependence}

As discussed before, in order to define bulk operators in a gauge invariant way, one can do it by throwing geodesics from the boundary. In this way, one can define the operator $\Phi(x,z)$ by throwing a spacelike geodesic from $x$ in the boundary up to some renormalized proper distance $z$. As discussed in \cite{Jafferis:2017tiu,Almheiri:2013hfa,error}, one can define the same operator $\Phi(x,z)$ for the family of states which contain the same geodesic. However, since $\Phi(x,z)$ will satisfy different wave equations in different backgrounds, when writing a boundary expression for this operator, it would seem to depend on the state. This is often called ``background dependence". 

Of course, whether two of these state independent (within the family of states with the same geodesic) operators are spacelike separated is going to depend on the state where the commutator is evaluated. Similarly, whether an operator $\Phi(x_R,z)$ is in the entanglement wedge of $R$ is state dependent. In this way, a gauge invariant description of the entanglement wedge in terms of boundary geodesics $(x_R,z)$ will be state dependent, so we don't have any reason to expect a state independent boundary representation for a bulk field (geodesically dressed to the boundary) in the entanglement wedge.   

However, it could be that there is a more state independent definition of bulk operators in the entanglement wedge which doesn't rely on throwing geodesics from the boundary. For example, one can define the state independent bulk operator integrated over the extremal surface anchored in $\partial R$. In any given state, this corresponds to a linear combination of our bulk zero modes\footnote{So that the smearing function in the RT surface doesn't depend on the state.} and, while in the boundary the zero modes depend on the modular hamiltonian explicitly, it is possible that this is just ``background dependence". This could be understood from the fact that we can think of background changing operator as as a bulk coherent state, so $|\psi'\rangle=U |\psi \rangle$. Since a bulk coherent state is linear in the fields, the unitary factorizes $U=U_r U_{\bar r}$ and the bulk modular hamiltonian changes by $K_{bulk,r}'=U_r K_{bulk,r} U_r^{\dagger}$. However, at this moment it is not clear to us if this argument is enough to show that these operators are state independent.

\section{Discussion}

In this paper, we have studied how to reconstruct bulk operators in the entanglement wedge in terms of boundary operators. Because of the equivalence between bulk and boundary modular flow, the natural boundary operators dual to bulk operators in the entanglement wedge have fixed modular frequency. The smearing function $f_w(X|x)$ is simply a convolution of the modular hamiltonian with the correlator between the bulk operator and a boundary operator with frequency $w$. This is well defined, although it might be hard to compute explicitly.  In the limit where the bulk operator goes to the RT surface, the expression simplifies and the smearing function is just the inverse of a bulk-to-boundary correlator.  

As we have explained before, in   \cite{Patricketal} a different connection between modular flow and entanglement wedge reconstruction was discussed. We would like to emphasize that while  eq. (7) of \cite{Patricketal} and our expression \nref{hkllmod} both have modular flow, these formulas are not related at all. The reason why there is modular flow in \nref{hkllmod} in the semiclassical limit where we are working (ie $G_N=0$), the bulk and boundary modular flows are equivalent and this provides a convenient way to parametrize operators. In contrast, \cite{Patricketal} use the modular hamiltonian to make the discussion of \cite{Dong:2016eik} stable for small but finite $G_N$, by smearing it with a fixed kernel in modular time, if one sets $G_N=0$ in their formula, one morally gets ${\cal O}_R=\text{tr}_{\bar R} \Phi(X_r)$.  

We would like to conclude with some further observations.

\subsubsection*{Bulk emergence}

It would be nice to understand to what extent one can use the expression for the zero modes \nref{zeromode} to probe the bulk locality directly from the boundary, by considering the zero modes of different regions. A key ingredient for bulk locality is the non-commutativity of $[\Pi(X),\Phi(X)]$, but the zero modes seem to be naturally probing $\Phi$ rather than $\Pi$. In other words, the equal time commutator of zero modes of different spatial regions (lying on the same spacelike slice in the bulk) is zero. Of course, $\partial_t O_0$ would naturally give us the $\Pi$ operator, but this doesn't seem a natural operator in $R$.  One could think that in principle, bulk locality can be studied by understanding the singularities of the correlators of zero modes \nref{G0eq} in the spirit of \cite{Heemskerk:2009pn,Maldacena:2015iua}.     Similar ideas have been pursued recently in   \cite{Kabat:2017mun,Sanches:2017xhn}. 

Note also that, even if modular flow is very complicated in general, \nref{Extra} implies that, in holographic theories, it is simpler than we would have expected, since it can be effectively described in terms of a single trace operator smeared over the whole boundary. 

\subsubsection*{Locality and the ``algebra'' of generalized free fields}

If we consider a given bulk $t=0$ slice, the usual HKLL discussion suggests that the algebra of bulk operators in that Cauchy slice is given in terms of $1/N$ corrected generalized free fields at the boundary $t=0$ slice 
combined with hamiltonian evolution\footnote{Since the HKLL operators only depend on the hamiltonian at one time \cite{HMPS}, we can restrict ourselves a time independent hamiltonian. }: $U_H=e^{i H t}$: $\lbrace \Phi(X,t=0) \rbrace = \lbrace {\cal O}(x,t=0), e^{i H t} \rbrace$. In this representation of the bulk hilbert space, if we divide $t=0$ into two subregions $R,\bar{R}$,  it is not clear how  ${\cal O}(x,t)$ factorizes into ${\cal O}_R \times {\cal O}_{\bar R}$. Even if it is clear that there should be such factorization at finite $N$, it doesn't seem like this factorization would have a smooth large $N$ limit: trying to take trace over $\bar{R}$ of a single trace operator (GFF) which has support across $\partial R$ seems rather ill defined. If one could do this, then by taking the trace over $R$ of the HKLL global representation of an operator in the entanglement wedge of $R$, one should get the identity in $\bar{R}$ (plus operators which annihilate any low energy state). 

What we have observed is that instead, we can think of the bulk operators at $t=0$ as being generated by boundary operators at $t=0$ and $U_K=e^{i (K_R-K_{\bar R}) t}$: $\lbrace \Phi(X,t=0) \rbrace = \lbrace {\cal O}(x,t=0), e^{i (K_R-K_{\bar R}) t} \rbrace$, for an arbitrary choice of $R$. For any given $R$, this representation of the bulk algebra explicitly preserves subregion locality.  

It is unclear to us how the algebra generated by the $1/N$ corrected generalized free fields and $U_K$ is (in the large $N$ sense) isomorphic to the algebra generated by the GFF's and $U_H$ (isomophic in the sense that the respectives GNS Hilbert spaces are the same). Generalized free fields at a given time can't be expanded in terms of generalized free fields at earlier times  because they don't satisfy any equation of motion (this is the reason why they violate the time slice axiom), but we don't need the whole $U_H$ for all times to get all linearly independent operators, somehow  $U_K$ is enough. This is reminiscent of the interesting idea \cite{Connes:1994hv} that modular time might play and interesting role in quantum gravity.

\subsubsection*{Quantum error correction }

In \cite{error,Dong:2016eik}, an expression for bulk reconstruction was given which depends on the Schmidt decomposition into $R,\bar{R}$ of states created by bulk excitations (see \cite{Patricketal} for a formula which generalizes the previous when the relative entropies are not exactly the same). It is not clear to what extent these formulas are different from  writing the global HKLL representation of the bulk operator and tracing over $\bar{R}$ , to get a localized operator in $R$ plus something that annihilates all simple states. And, as discussed in the previous few paragraphs, we don't expect this to have a smooth large $N$ description. 

Our discussion didn't use any of the insights from quantum error correction, we have just followed the usual HKLL approach with the knowledge that the bulk and boundary modular flows are equivalent.  However, if we wanted to use the formulas of \cite{error,Dong:2016eik} in this bulk perturbative setting, their interpretation would be that they make the extrapolate dictionary more precise. That is, we would first write the bulk fields in a basis labeled by a position in the boundary  $AdS$ and then map them to boundary operators where it is now clear how to take the trace. Given that the bulk and boundary operators related by the extrapolate dictionary live in different Hilbert spaces, one has to be careful when taking the trace over the complement of the region and this is what the formulas in the appendix of \cite{error} achieve in this setting: they map the bulk operator $\Phi_r(x,z=0)$ to the boundary operator ${\cal O}_R(x)$.

Finally, note that in order to write the bulk operators in a basis where it is easy to take the boundary trace, we had to use an insight of bulk low energy subspaces: the modular flow of a simple operator is a low energy operator and thus acts within the code subspace (see footnote $5$). We don't expect this to be a generic property of quantum error correcting codes.

\section*{Acknowledgements}

 We would like to thank Ahmed Almeihri, Horacio Casini, Bartek Czech, Xi Dong, Patrick Hayden, Michal Heller, Daniel Jafferis,  Lampros Lamprou, Juan Maldacena, Don Marolf, Sam McCandlish, Onkar Parrikar,  Mukund Rangamani, Michael Walter and Tianci Zhou for discussions. We would also like to thank the organizers of the Solvay Workshop on Holography for Black Holes and Cosmology, where this project was started. AL acknowledges support from the Simons Foundation through the “It from Qubit” collaboration. AL would also like to thank the Department of Physics and Astronomy at the University of Pennsylvania for hospitality during the devolpment of this work. TF is supported in part by a DARPA YFA, contract D15AP00108.
 \newpage
\appendix
\section{Zero modes correlators for local modular hamiltonians} \label{AA}

 In this appendix, we are going to show explicitly for a local modular hamiltonian that :
 \begin{equation}
 G_0(Y_1,Y_2)=\int_{-\infty}^{\infty} ds \langle {\cal O}_s(Y_1) {\cal O}(Y_2) \rangle= 4\pi \int_{H_{d-1}} dY_{B} \langle \Phi(Y_B) {\cal O}(Y_1) \rangle \langle \Phi(Y_B) {\cal O}(Y_2) \rangle 
 \end{equation}
 
 This result is just a simple modification of the discussion of \cite{Faulkner:2014jva}. 
 
 First, if we write the boundary correlators in hyperbolic space (and equal times), we have that:
 \begin{equation}
 G_0(Y_1,Y_2)=\int_{0}^{\infty} \frac{d\lambda}{\lambda} \frac{c_{\Delta}}{ (-2 Y_1 \cdot Y_2+\lambda+\lambda^{-1})^{\Delta}}; c_{\Delta}=\frac{(2\Delta-d) \Gamma(\Delta)}{ \pi^{d/2} \Gamma(\Delta-d/2)}
 \end{equation}
 
 We can exponentiate the $Y_1 \cdot Y_2
 $ dependence by introducing a Schwinger parameter:
 \begin{equation}
  G_0(Y_1,Y_2)=\frac{c_{\Delta}}{\Gamma(\Delta)}\int_{0}^{\infty} \frac{d\lambda}{\lambda} \int_{0}^{\infty} \frac{d t}{ t} t^{\Delta} \exp(-t (\lambda+\lambda^{-1}-2 Y_1 \cdot Y_2))
 \end{equation}
 
 We can change integration variables to $(t_1,t_2)=(\sqrt{t \lambda},\sqrt{t}/\sqrt{\lambda})$, which let us write the exponential as $-(t_1 Y_1+t_2 Y_2)^2$. 
  \begin{equation}
  G_0(Y_1,Y_2)=\frac{2 c_{\Delta}}{\Gamma(\Delta)}\int_{0}^{\infty} \frac{d t_1}{t_1} \int_{0}^{\infty} \frac{d t_2}{ t_2} (t_1 t_2)^{\Delta} \exp(-(t_1 Y_1+t_2 Y_2)^2)
 \end{equation}
 
 As shown in \cite{Faulkner:2014jva}, inside the $t_1,t_2$ integral, one can substitute:
 \begin{equation}
  \exp(-(t_1 Y_1+t_2 Y_2)^2) \rightarrow \frac{2 \pi c_{\Delta}}{ \Gamma(\Delta) (2 \Delta-d)^2} \int_{H_{d-1}} dY_B e^{2 Y_B\cdot (t_1 Y_1+t_2 Y_2)}
 \end{equation}
 
 This makes the $t_1,t_2$ integrations independent and simple and thus one just gets:
 \begin{equation}
 G_0(Y_1,Y_2)=4\pi  \int_{H_{d-1}} dY_B \frac{c_{\Delta}}{(2 \Delta-d) (-2 Y_B \cdot Y_1)^{\Delta}}\frac{c_{\Delta}}{(2 \Delta-d) (-2 Y_B \cdot Y_2)^{\Delta}}
 \end{equation}
 which is the bulk equation for the zero modes, given that:
 \begin{equation}
 \langle \Phi(Y_B) {\cal O}(Y_1) \rangle=\frac{c_{\Delta}}{2\Delta-d} \frac{1}{(-2 Y_B.Y_1)^{\Delta}}
 \end{equation}

 \section{Alternative derivation of the zero mode formula}

We start this appendix with a slight detour to discuss the behavior of modular flow for local modular Hamiltonians and then discuss to what extent we can extrapolate the action of local modular flow to fields near the entanglement cut where locally the cut looks like a Rindler cut.
 
\subsection{Local modular hamiltonians}

For certain symmetric cases, the modular hamiltonian is local: the flow is geometric with respect to a killing vector $\xi(x)$ :
\begin{equation}
[K, {\cal O}(x)]=\xi^{I}(x) \partial_I {\cal O}(x)
\end{equation}
That is, ${\cal O}(x,s)={\cal O}(x_{\xi}(s))$ and the modular hamiltonian in position space will simply be the integral of the stress tensor. This means that we can write $K=\int_R d\Sigma^{\mu} \xi^{\nu} T_{\mu \nu}$.

\subsubsection*{Ultralocality}

Consider the half space/Rindler cut in vacuum. The entangling surface in lightcone coordinates  $d u d v+dy^2$ corresponds to $u=v=0$. We can consider the horizon $u=0$ as the Cauchy slice where we consider our free fields. The Rindler cut corresponds to the future horizon $ v >0$.   
Fields in the future horizon are ultralocal: the algebra of operators is a direct product of the two dimensional algebra of operators at every $(y,v)$, that is correlators between different lightrays vanish:

\begin{equation}
\langle \partial_{v} \Phi(v,y) \Phi(v',y') \rangle=\frac{1}{4\pi(v-v')} \delta(y-y')
\end{equation}

In this case, modular evolution just amounts to rescaling $v \rightarrow v e^{s}$. Given that correlators in the vacuum state are ultralocal, the modular hamiltonian acts independently on each light ray. Locality of the modular hamiltonian and ultralocality are equivalent statements: if a state has ultralocal correlators, then the modular hamiltonian necessarily acts independently in each lightray. If the modular hamiltonian is local, given that the "bilocal" kernel in modular Fourier space defined in \eqref{bulkKw} is the inverse of the commutator $G_w$, the commutator has to be ultralocal which then implies the correlator is ultralocal.

The algebra of operators in each lightray $(y,u)$ is basically that of the (chiral half of a) $2 d$ massless scalar CFT. So for example the correlator $ \langle \Phi(v,y) \Phi(0,y) \rangle \sim (4\pi)^{-1} \log(v) \delta(y-y') $ has a strong IR divergence which we would usually try to avoid. In the next section these $\ln$ divergences naturally arise for more general entangling cuts due to the approximate locality of modular flow close to the cut.

\subsubsection*{Locality of the modular flow close to the entangling surface}

One would like to understand to what extent the modular flow is local close to more general entangling surfaces.  For free fields, one can explore this by studying the extent to which correlators are ultralocality in this more general case. This is because for free fields (and a gaussian state satisfying Wick's theorem) knowledge of the two point correllator of a state is enough to reproduce the modular Hamiltonian.

For general surfaces , 
it will  be convenient to write the free fields in terms of  adapted lightfront coordinates. For a given codimension two surface we will denote coordinates along this surface as $y$. By shooting lightrays from this surface, we can define the future/past horizons ${\cal H}^{\pm}$. We can introduce coordinates $u,v$ which denote some fixed affine distance along these lightrays. In general, the future/past horizons will have caustics, so these coordinates will breakdown at some finite affine parameter. We would always like to consider a family of space-like Cauchy slices which limits to the horizon ${\cal H}^+$ - at least for some portion of ${\cal H}^+$ avoiding the caustics. So in the limit we will choose to parameterize our Cauchy slice with the null coordinate $v$ along ${\cal H}^{+}$ for $0< v < v_\star$ after which $v$ becomes a space-like coordinate and the Cauchy slice moves into the interior of the domain of dependence of the entanglement region. We will denote this smoothed out Cauchy-slice as ${\cal H}^{+}_{smooth}$.

For $u\sim 0$, $v \ll 1$, we can use adapted coordinates to write the metric of this surface:
\begin{equation}
ds^2=du dv+(\gamma_{i j}+K^{+}_{i j} v)dy_i dy_j+O(u)+O(v^2)
\end{equation}
The presence of this extrinsic curvature implies that for neighbourging light rays ($y \sim y'$), ultralocality will be preserved as long as $v K^+ \ll 1$. This can be seen explicitly from the correlator:
\begin{equation}
\label{locmod}
\langle \partial_{v} \Phi(v,y) \Phi(v',y') \rangle \mathop{=}_{y \sim y'}\frac{1}{4\pi(v-v')} \delta(y-y')+O(v K^+,v' K^+)
\end{equation}
This quantifies what one means by the modular hamiltonian being approximately local - it will be approximately local if we act on operators with $v  \ll 1/K^+$. Thus we can approximate the flow via $v \rightarrow v e^s$ in this case.
 
Note that for $y$ far from $y'$ ultralocality will certainly break down since and the non-locality of the modular hamiltonian will depend on more global geometric properties. For example, if we have two disconnected surfaces with $K_{i j}^{+}=0$, it is clear that the correlators between two lightrays  emanating from the different surfaces at $y,y'$ won't be ultralocal, rather $\langle \partial_{v} \Phi(v,y) \Phi(v',y') \rangle \rightarrow C(y,y')$ for small $v,v'$.
However this correlator will be suppresed in the distance between the lightrays and comparing to \eqref{locmod} we expect modular flow to produce non-local mixing with distant light rays, but this effect is supressed by the correlator $C(y,y')$ between these two distant points and positive powers of $v,v'$.

To sum up, for general surfaces, ${\cal H}^{+}$ has to be smoothed out because it contains caustics. Generally the presence of caustics spoils ultralocality and thus the locality of the modular hamiltonian. However, for fields close to the entangling surface, we expect that as long as $v$ is smaller than any of the other length scales of the system, we still expect the modular hamiltonian to be local.

While this analysis seemed to rely explicitly on the ultralocality of free fields, the lesson is more general: we expect the modular flow to act locally close to the boundary of a region, as long as we don't probe any other scale of the system. One can argue for this by considering the expectation value of operators in the presence of conical singularity of strength $\frac{2 \pi}{n}$ (which would compute the Renyi entropies), inserted along the entangling surface. In conformal field theories, if we consider a field close to the conical singularity (denote $\rho$ the polar distance to the singularity),  we expect:

\begin{equation}
\langle {\cal O}_{\Delta}\rangle_n=c_{\cal O}(n) {\rho}^{-\Delta}+O(\rho / L)
\end{equation}
That is, we expect these one point functions to diverge with their anomalous dimension as they get close to the singularity, and for a general state and region, this UV conformal behaviour will be leading as long as we consider $\rho \ll L$, where $L$ is the shortest length in the system. Upon analytic continuation in $n$, we expect that this notion of approximate locality  also applies to modular evolution.

 \subsection{Zero mode from local modular flow \label{AB}} 
 
Consider the object
\be
 \int_{-\infty}^\infty d s e^{ is \omega} \left< \left[ \Phi_s(X),\mathcal{O}(x) \right] \right> 
\ee
in the limit where the bulk operator goes to the RT surface,  $X \rightarrow X_{RT}$. In this limit the commutator $\left< \left[ \Phi_s(X),y \right] \right>$ becomes a constant for a long period of modular time $ - \log \rho < s < \log \rho$, where $\rho$ is the approximate Rindler radial coordinate, the constant
is $0$ since the RT surface and the boundary point are space-like separated (given than the modular flow is local as long as $\rho$ is smaller than any other scale). For most
of this discussion we assume the fourier transform is well defined, so the commutator vanishes
sufficiently fast for large $s$.  

The correlator similarly is constant for a long period of modular time $ - \log \rho < s < \log \rho$
and the fourier transform evaluates to a resolved delta function:
\be
\int ds e^{- i s \omega} \left< \Phi_s(X) \mathcal{O}(x) \right>\approx  \delta_{(\log \rho)^{-1}} ( \omega)  \left< \Phi(X_H) \mathcal{O}(x) \right> 
\ee
where  the resolution scale is related to $\log(\rho)$ and in particular $ \delta_{\log \rho^{-1} } (0) \approx 2 \log \rho $.

In this way,  in the limit where the bulk operator goes to the RT surface, \nref{smearing} reduces to

\be
\Phi(X_{RT}) = \int_R dx F(X_{RT} | x) \mathcal{O}_0(x) \label{RTsmear}
\,, \qquad F(X_{RT}|x) = \int d y \mathcal{K}_0(x,y) \left< \Phi(X_{RT}) \mathcal{O}(y) \right>
\ee
 Now the idea
is to take correlators of \eqref{RTsmear} with respect to $\Phi(Y)$ for $Y$ close to the entangling surface.
Doing this carefully and using the same as above to take $Y \rightarrow Y_{RT}$ we have:
\be
\left< \Phi(X_{RT}) \Phi(Y) \right> \mathop{=}_{Y \approx Y_{RT}} 
 \delta_{(\log \rho)^{-1} } (0) \int d x d y F(X_{RT}|x) \left<\mathcal{O}(y) \Phi(Y_{RT}) \right>
\ee
The $\log \rho$ divergence in the right hand side above could seem surprising at first. Of course, as discussed in the previous section, we expect this logarithmically divergent term from the short distance behaviour of the correlator:
\be
\left< \Phi(X_{RT}) \Phi(Y) \right> \mathop{=}_{Y \approx Y_{RT}; X_{RT} \approx Y_{RT}} 
\frac{N_d}{ ( (x_{RT} - y_{RT})^2 + \rho^2 )^{\frac{d-2}{2}} } 
\ee
where $\rho$ is a regulator which separates the two operators away from the RT surface
and $x_{RT}$ and $y_{RT}$ are $d-2$ flat coordinates along the RT surface in a patch around $X_{RT}$ and $Y_{RT}$. One can work out that $N_d^{-1} =2 \pi S_{d-3}$ where $S_{d-3}$ is the area of a $d-3$ sphere. 
As a distribution this behaves as a delta function:
\be
\left< \Phi(X_{RT}) \Phi(Y) \right>   \mathop{=}_{Y \approx Y_{RT}}
\frac{1}{2 \pi} \log  (\frac{\rho}{L}) \delta^{(d-2)}(X_{RT},Y_{RT})  + (\rm{finite ~ as ~ } \rho \rightarrow 0) 
\ee
where $L$ is some IR scale associated to the curvature scale or the mass of the scalar and these formulas are consistent with our previous discussion around \eqref{locmod} if we set
$\rho = \sqrt{ uv }$.

Comparing the $\log \rho$ divergent terms we see that:
\be
\delta(X_{RT},Y_{RT}) = 4\pi \int d x  F(X_{RT}|y) \left<\mathcal{O}(x) \Phi(Y_{RT}) \right>
\ee
Let's write this equation as $1 =4\pi F \cdot \Phi$ where integration has become
a matrix product and we assume all these matrices are invertible (both left and right inverse exist), which means that $F^{-1}=4\pi \Phi$. 
Likely these are anyway only formal manipulations since we already know that
$F$ is only defined as a distribution acting on CFT correlation functions.  The reconstruction of operators on the RT surface is defined as $\Phi = F \cdot \mathcal{O}_0$ so that we can write $4 \pi \Phi \cdot \Phi = \mathcal{O}_0$, which when written out looks like:
\be
4\pi \int d X_{RT} \left< \mathcal{O} (x) \Phi(X_{RT}) \right> \Phi(X_{RT}) =
\mathcal{O}_0(x) 
\ee
which is the same formula as we have derived in the main text.

\end{document}